# Configurational Temperature Control for Atomic and Molecular Systems

*Karl P. Travis** and *Carlos Braga*

Immobilisation Science Laboratory, Department of Engineering Materials, University of Sheffield, Mappin Street, Sheffield S1 3JD, UK.

k.travis@sheffield.ac.uk

ABSTRACT.

A new configurational temperature thermostat suitable for molecules with holonomic constraints is derived. This thermostat has a simple set of motion equations, can generate the canonical ensemble in both position and momentum space, acts homogeneously through the spatial coordinates, and does not intrinsically violate the constraints. Our new configurational thermostat is closely related to the kinetic temperature Nosé-Hoover thermostat with feedback coupled to the position variables via a term proportional to the net *molecular* force. We validate the thermostat by comparing equilibrium static and dynamic quantities for a fluid of *n*-decane molecules under configurational and kinetic temperature control. Practical aspects concerning the implementation of the new thermostat in a molecular dynamics code and the potential applications are discussed.

1. **Introduction**

The ability to control temperature in a simulation is important in many situations such as the study of phase equilibria, becoming essential for non-equilibrium simulations, where a non-equilibrium steady state may not be obtained without it. Temperature may be controlled by a number of different methods ranging from stochastic control to deterministic thermostats that appear in the equations of motion. Examples of the latter include the Gaussian thermostat[1], the Nosé-Hoover thermostat[2], and Dettmann and Morriss' reformulation of the Nosé-Hoover thermostat based on a vanishing Hamiltonian[3] (later rediscovered as the Nosé-Poincaré thermostat by Bond *et al.*[4]). Until fairly recently, each of these thermostats was employed to maintain the so-called kinetic temperature. That is, the temperature defined by classical equipartition theory through an ensemble average of the kinetic energy of the constituent particles or via the time average of the instantaneous kinetic energy. This temperature is straightforward to define and poses few difficulties when controlled by a thermostat in an equilibrium simulation. However, in systems which are driven away from equilibrium such as in a simulation of Couette flow, it is the thermal (peculiar) kinetic energy which must be controlled, and this is not known accurately *a priori*.

The introduction of a tractable statistical mechanical expression for the thermodynamic temperature by Rugh was an advance in the field of temperature control[5]. Using a similar approach to Rugh, Butler *et al* developed a configurational temperature expression[6]. This definition of temperature, as the name implies, is calculated using only configurational information. It is interesting to note that this was not the first appearance of a mean square force definition of temperature[7]; Butler *et al*'s expression also appears in the 1952 Russian Language edition of Landau and Lifshitz's textbook,*Statistical Physics*[8]. The notion of configurational temperature may even predate this work; the discussion concerning generalized equipartition in Tolman's textbook, *The principles of Statistical Mechanics* [9] first published in 1938, alludes to its existence without ever stating it explicitly. The full importance of this new (or rediscovered) configurational temperature was not realized at first; it was proposed as a diagnostic tool for checking Monte Carlo codes by Butler and co-workers.

The significance of configurational temperature became apparent after the introduction of the first configurational thermostat by Delhommelle and Evans (DE)[10]. Configurational temperature control is important because it avoids the necessity of accurately knowing the local streaming velocity in an NEMD simulation. DE demonstrated that string phases were absent when a system of LJ atoms were subjected to planar Couette flow at high shear rates with configurational temperature control.

Applying thermostats to molecular systems introduces a new dilemma to the simulator: which degrees of freedom to thermostat. A simulation of rigid bodies may employ either rotational or translational temperature control. For flexible molecules, in which the equations of motion are solved for the atomic sites along with either intramolecular forces or holonomic constraints to keep the atoms in the form of a molecule, there are several choices of temperature control. The translational temperature of the centre-of-mass motion may be controlled, as may the total kinetic energy (and hence, atomic temperature). Such a choice may no longer be one of convenience. If the simulation involves large molecules such as a protein or polymer, thermostatting just three translational degrees of freedom out of a great many molecular degrees of freedom may result in a slower equilibration rate as it takes time for the extremities of the molecule to reach thermal equilibrium with the centre-of-mass. Away from equilibrium the situation is more complicated than for atoms. Now one is faced with the need to know *a priori* either the streaming velocity at an atomic site or that of the centre-of-mass. The streaming velocity at an atomic site will have contributions from both the translational and rotational motions of the molecule. For rigid uniaxial molecules it is possible to develop a range of profile unbiased thermostats (PUTs) which can be used in NEMD simulations that are free from artefacts such as string phases and antisymmetric stress [11, 12]. However, there is no satisfactory PUT for the general case.

The configurational temperature concept can also be applied to molecular systems. Lue and Evans (LE) developed Rugh's original ideas, first defining a configurational molecular temperature, then extending the DE atomic thermostat to deal with molecules which may contain constraints[13]. However, the LE thermostat, like the atomic DE thermostat, involved adding a term proportional to the gradient of the configurational temperature to the equations of motion for the atomic positions. Since the temperature gradient will be different for each site, any bond constraints that may be in operation are instantaneously violated by this thermostat. Lue and Evans solved this constraint violation problem by adding an extra term to the position equation of motion to counteract this effect[13]. Unfortunately this makes the implementation of the LE thermostat even more cumbersome than the DE thermostat. Furthermore, it is not clear that the LE equations can generate the canonical distribution of positions.

We have recently improved upon the DE atomic thermostat by developing two Nosé-Hoover style configurational thermostats which were designed so that they would generate the canonical ensemble distribution in position and momentum space[14]. This was achieved by finding a set of motion equations that satisfied a generalized Liouville equation. The equations of motion for either of these new thermostats contain a term proportional to the force acting on an atom.

In this article we develop a pair of configurational molecular thermostats in the same manner as our earlier atomic thermostats. These new thermostats have particularly simple equations of motion; the extra term involves the force acting on a molecule and thus there is no requirement to add correction terms for constraint violation since they are always satisfied by definition. We validate the new thermostats by comparison with the kinetic thermostats for a fluid of *n*-decane molecules.

## 2. Microscopic Expressions for Temperature

The thermodynamic temperature, $T$, is defined through the classical thermodynamic derivative

$$\frac{1}{T} = \left.\frac{\partial S}{\partial E}\right|_V, \qquad (1)$$

where $S$ is the entropy, $E$ is the internal energy and the derivative is taken at fixed volume. In statistical mechanics, the entropy of an isolated Hamiltonian system with total energy, $E$, is given by

$$S(E) = k_B \ln \int_{\mu C(E)} d\boldsymbol{\Gamma} \tag{2}$$

where $\boldsymbol{\Gamma}$ is the usual phase space vector and $\mu C(E)$ indicates that the integration is restricted to regions in phase space where the internal energy lies within the bounds $E < H(\boldsymbol{\Gamma}) < E+\delta E$ where $\delta E << E$.

Rugh obtained a dynamical expression for temperature from a geometric analysis of phase space[5]. He showed that such a measure of temperature is related to the curvature of an energy surface in the microcanonical ensemble. The proof involves concepts in measure theory and we do not repeat it here. The interested reader may consult Jepps *et al* [15] or Jepps[16] for alternative (and generalized) versions of the proof as well as the second paper by Rugh on this topic[17]. Rugh was able to show that the thermodynamic temperature could be obtained via a time average of the functional, $\nabla \cdot \left( \nabla H / \|\nabla H\|^2 \right)$, on an energy surface, where $\| \cdot \|$ denotes a norm, and $H$ is the system Hamiltonian, giving

$$\frac{1}{k_B T} = \left\langle \nabla \cdot \frac{\nabla H(\boldsymbol{\Gamma})}{\|\nabla H(\boldsymbol{\Gamma})\|^2} \right\rangle \tag{3}$$

Rugh later generalized his derivation to yield a new expression for temperature, still valid in the microcanonical ensemble, but now given in terms of a family of vector fields, $\boldsymbol{B}(\boldsymbol{\Gamma})$[16]

$$\frac{1}{k_B T} = \left\langle \nabla \cdot \frac{\boldsymbol{B}(\boldsymbol{\Gamma})}{\nabla H(\boldsymbol{\Gamma}) \cdot \boldsymbol{B}(\boldsymbol{\Gamma})} \right\rangle \tag{4}$$

This expression was also simultaneously and independently derived (using different methods) by Jepps *et al*[15] and Rickayzen and Powles[18].

An alternative expression for temperature valid in both the canonical and microcanonical ensembles, was derived by Jepps *et al*[15],

$$\frac{1}{k_B T} = \frac{\langle \nabla \cdot \boldsymbol{B}(\boldsymbol{\Gamma}) \rangle}{\langle \nabla H(\boldsymbol{\Gamma}) \cdot \boldsymbol{B}(\boldsymbol{\Gamma}) \rangle}, \tag{5}$$

which is distinct from either of Rugh's temperature expressions, being defined as the fraction of two average quantities. The vector field appearing in eqs. 4-5, $\boldsymbol{B}(\boldsymbol{\Gamma})$, while fairly arbitrary, must satisfy a number of criteria, namely that $0 < |\langle \nabla H \cdot \boldsymbol{B}(\boldsymbol{\Gamma}) \rangle| < \infty$, $0 < |\langle \nabla \cdot \boldsymbol{B}(\boldsymbol{\Gamma}) \rangle| < \infty$ and that $\langle \nabla \cdot \boldsymbol{B}(\boldsymbol{\Gamma}) \rangle$ should grow more slowly than $e^N$ in the thermodynamic limit. The first two of these criteria result from the use of Gauss' theorem in the proof by Jepps *et al*. Jepps *et al* point out that eq. 5 appears without proof in the well-known textbook by Allen and Tildesley[19], which would suggest it is another example of a re-discovered equation. However, this would miss the point; without the rigorous proofs provided by Rugh and Jepps, the conditions under which the configurational temperature is properly defined, and the restrictions placed on the vector field, $\boldsymbol{B}$, would not be apparent.

If the choice of $\boldsymbol{B}(\boldsymbol{\Gamma})$ is limited to those vector fields which are linear transformations of $\nabla H$ ie.

$$\boldsymbol{B}(\boldsymbol{\Gamma}) = \frac{X(\boldsymbol{\Gamma})}{\nabla H \cdot X(\boldsymbol{\Gamma})} \tag{6}$$

where $X(\boldsymbol{\Gamma})$ is a different arbitrary vector field, then $\nabla H(\boldsymbol{\Gamma}) \cdot \boldsymbol{B}(\boldsymbol{\Gamma})$ is guaranteed to be unity for all choices of $X(\boldsymbol{\Gamma})$, and one recovers Rugh's generalized expression for temperature (eq. 4). Taking $X(\boldsymbol{\Gamma}) = \nabla H$ recovers eq. 3. Alternatively, substituting $X(\boldsymbol{\Gamma}) = \boldsymbol{G} \cdot \nabla H(\boldsymbol{\Gamma})$ in eq. 5, where $\boldsymbol{G}$ is a general metric matrix, gives

$$\frac{1}{k_B T} = \left\langle \nabla \cdot \frac{\bm{G} \cdot \nabla H}{\nabla H \cdot \bm{G} \cdot \nabla H} \right\rangle. \tag{7}$$

The form of $\bm{G}$ can be fairly arbitrary as long as the condition, $\nabla H(\bm{\Gamma}) \cdot \bm{G} \cdot \nabla H(\bm{\Gamma}) \neq 0$, is satisfied. By carefully choosing $\bm{G}$, one can derive different temperature expressions, including configurational temperatures [15].

Alternatively, taking $\bm{B}(\bm{\Gamma}) = \bm{G} \cdot \nabla H(\bm{\Gamma})$ and using eq. 5 gives

$$\frac{1}{k_B T} = \frac{\langle \nabla \cdot \bm{G} \cdot \nabla H \rangle}{\langle \nabla H \cdot \bm{G} \cdot \nabla H \rangle} \tag{8}$$

The above generalization of Rugh's expression to arbitrary vector fields and different ensembles has the distinct disadvantage that it permits an infinite array of possible definitions of temperature without yielding a method for selecting the optimum choice of vector field. Clearly, more research is necessary in this area to help narrow down the choice. In the absence of any new theoretical developments, and faced with ambiguity in alternative definitions of temperature one should select a definition of temperature based on the following advice[20]: (i) it should have smaller fluctuations, (ii) lower powers of $p$ or $q$, and (iii) have simpler physics.

### 2.1 Temperature expressions for systems with holonomic constraints

The various expressions for thermodynamic temperature defined by eqs. 3-8, are formally the same for atomic and molecular systems. However, it is frequently the case that molecules are modeled with some degrees of freedom removed through the use of holonomic constraints. These are usually bond distance constraints and sometimes second nearest neighbor distance constraints (fixed valence angles). In these cases the arbitrary vector field, $\bm{B}(\bm{\Gamma}) = \bm{G} \cdot \nabla H(\bm{\Gamma})$, must be orthogonal to the hypersurface of constraint[16].

Consider a 3-dimensional system (extension to $d$ dimensions is trivial) of $N$ molecules, with each molecule $i$ containing $n_i$ atoms or sites. The phase space vector for such a system can conveniently be written in component form as $\bm{\Gamma} = \{\Gamma_{i\alpha'a}\}$, where $a$ is either a position coordinate $(a = 1,2,3)$ or a momentum coordinate $(a = 4,5,6)$ of site $\alpha'$ in molecule $i$ [21]. The temperature expression eq. 7 for this system may be given explicitly by

$$\frac{1}{k_B T} = \sum_{i\alpha'a} \sum_{j\alpha''b} \left\langle \partial_{i,\alpha',a} \left\{ \frac{g_{i\alpha'a,j\alpha''b}[\partial_{j\alpha''b} H(\Gamma)]}{\sum_{i,\alpha',a} \sum_{j,\alpha'',b} [\partial_{i\alpha'a} H(\Gamma)] g_{i\alpha'a,j\alpha''b} [\partial_{j\alpha''b} H(\Gamma)]} \right\} \right\rangle \quad (9)$$

where $\partial_{i\alpha'a}$ is shorthand for the differential operator, $\partial/\partial \Gamma_{i\alpha'a}$, and $g_{i\alpha'a,j\alpha''b}$ are the matrix elements of **G**. Alternatively, if the temperature expression denoted by eq. 8 is used, we have

$$\frac{1}{k_B T} = \frac{\left\langle \sum_{i\alpha'a} \partial_{i\alpha'a} \sum_{j\alpha''b} g_{i\alpha'a,j\alpha''b}[\partial_{j\alpha''b} H(\Gamma)] \right\rangle}{\left\langle \sum_{i\alpha'a} \sum_{j\alpha''b} [\partial_{i\alpha'a} H(\Gamma)] g_{i\alpha'a,j\alpha''b} [\partial_{j\alpha''b} H(\Gamma)] \right\rangle} \quad (10)$$

If the above molecular system is subject to a set of *K* holonomic constraints of the general form

$$\sigma_\alpha(\Gamma) = 0 \quad \text{for} \quad \alpha = 1,2\text{K } K. \quad (11)$$

then it can be shown[21] that the matrix **G** must be chosen so that

$$\sum_{i\alpha'a} \sum_{j\alpha''b} [\partial_{i\alpha'a} H(\Gamma)] g_{i\alpha'a,j\alpha''b}(\Gamma) [\partial_{j\alpha''b} \sigma_\alpha(\Gamma)] = 0 \quad \text{for} \quad \alpha = 1,2\text{K},K. \quad (12)$$

This condition arises from the need to ensure that the phase points (which are displaced in a direction parallel to the energy gradient) still satisfy the constraints at their new location.

A set of *K* distance constraints take the following form

$$\sigma_\alpha(\mathbf{\Gamma}) = (\mathbf{r}_{i\alpha'} - \mathbf{r}_{i\alpha''})^2 - d_\alpha^2 = 0 \quad \text{for} \quad \alpha = 1, 2\text{K}, K. \tag{13a}$$

$$\sigma_\alpha(\mathbf{\Gamma}) = (\mathbf{r}_{i\alpha'} - \mathbf{r}_{i\alpha''}) \cdot (\dot{\mathbf{r}}_{i\alpha'} - \dot{\mathbf{r}}_{i\alpha''}) = 0 \quad \text{for} \quad \alpha = K+1, K+2\text{K}, 2K, \tag{13b}$$

where $d_\alpha$ is the desired length of each bond and $\dot{\mathbf{r}}_{i\alpha'}$ are the Cartesian velocities of the atomic sites. The first set of constraints ensures that the bonds are maintained at fixed distances, while the second set result from the fact that the momenta conjugate to the constrained degrees of freedom must vanish (13b is the time derivative of eq. 13a).

If the matrix elements are chosen such that

$$g_{i\alpha'a,j\alpha''b}(\mathbf{\Gamma}) = \begin{cases} \delta_{ij}\delta_{ab} & \text{if } a \text{ and } b \leq 3 \\ 0 & \text{otherwise,} \end{cases} \tag{14}$$

then eq. 12 is guaranteed to be satisfied, and from eq. 7, one recovers an expression for an order 1 molecular version of $T_{con1}$,

$$k_B T_{con1} = \left\langle \frac{\sum_{i=1}^{N} \sum_{\alpha',\alpha''=1}^{n_i} \frac{\partial \Phi}{\partial \mathbf{r}_{i\alpha'}} \cdot \frac{\partial \Phi}{\partial \mathbf{r}_{i\alpha''}}}{\sum_{i=1}^{N} \sum_{\alpha',\alpha''=1}^{n_i} \frac{\partial}{\partial \mathbf{r}_{i\alpha'}} \cdot \frac{\partial \Phi}{\partial \mathbf{r}_{i\alpha''}}} \right\rangle + O(1/N) \tag{15}$$

which was derived by Lue and Evans[21] and in which $\Phi$ is the potential energy. The same choice of matrix elements used in the definition of temperature defined by eq. 10 gives rise to a second expression for the molecular configurational temperature, which we shall denote as $T_{conF}$,

$$k_B T_{conF} = \frac{\left\langle \sum_{i=1}^{N} \sum_{\alpha',\alpha''=1}^{n_i} \frac{\partial \Phi}{\partial \mathbf{r}_{i\alpha'}} \cdot \frac{\partial \Phi}{\partial \mathbf{r}_{i\alpha''}} \right\rangle}{\left\langle \sum_{i=1}^{N} \sum_{\alpha',\alpha''=1}^{n_i} \frac{\partial}{\partial \mathbf{r}_{i\alpha'}} \cdot \frac{\partial \Phi}{\partial \mathbf{r}_{i\alpha''}} \right\rangle} \tag{16}$$

Kinetic expressions for the temperature of a molecular system are well known from Equipartition Theory. However, it is an interesting exercise to derive kinetic based temperatures from Rugh and Jepps' treatments. First we note that if we have the following choice of the metric matrix, $G$,

$$g_{i\alpha'a, j\alpha''b}(\boldsymbol{\Gamma}) = \begin{cases} \dfrac{m_{i\alpha'} m_{i\alpha''}}{M_i} \delta_{ij}\delta_{ab} & \text{if } a \text{ and } b \leq 3 \\ 0 & \text{otherwise,} \end{cases} \quad (17)$$

we can generate the following expression for the (molecular) kinetic temperature (sometimes referred to as the translational temperature) upon substitution of eq. 17 into eq. 10:

$$\frac{1}{k_B T_{kin}} = \frac{\left\langle \sum_{i\alpha'} \dfrac{m_{i\alpha'}}{M_i} \dfrac{\partial}{\partial \boldsymbol{p}_{i\alpha'}} \cdot \sum_{\alpha''} \boldsymbol{p}_{i\alpha''} \right\rangle}{\left\langle \sum_i \dfrac{\boldsymbol{p}_i \cdot \boldsymbol{p}_i}{M_i} \right\rangle} \quad (18)$$

At first sight eq. 18 appears to miscount the number of degrees of freedom. However, it is important to realize that the set of molecular momenta are not all independent. Momentum conservation implies that the following relations hold,

$$\sum_i \sum_{\alpha'} \boldsymbol{p}_{i\alpha'} \equiv \sum_i \boldsymbol{p}_i = \boldsymbol{C} \quad (19)$$

where $\boldsymbol{C}$ is a constant vector which is usually chosen to be zero. The differentiation in the numerator of eq. 18 must therefore be carried out on a set of independent momenta. This can be achieved by re-writing any one of the molecular momenta in terms of the ($N$-1) remaining ones. The result of this manipulation is to reduce the number of terms in the outer summation by one, giving

$$\sum_{i\alpha'} \frac{m_{i\alpha'}}{M_i} \frac{\partial}{\partial \boldsymbol{p}_{i\alpha'}} \cdot \sum_{\alpha''} \boldsymbol{p}_{i\alpha''} = \sum_{i}^{N-1} \sum_{\alpha'} \frac{m_{i\alpha'}}{M_i} \frac{\partial}{\partial \boldsymbol{p}_{i\alpha'}} \cdot \sum_{\alpha''} \boldsymbol{p}_{i\alpha''} = 3(N-1) \tag{20}$$

which, when substituted back into eq. 18 yields the well known expression for translational kinetic temperature.

The following judicious choice for the matrix elements,

$$g_{i\alpha'a, j\alpha''b}(\boldsymbol{\Gamma}) = \begin{cases} m_{i\alpha'} \delta_{ij} \delta_{\alpha'\alpha''} \delta_{ab} & \text{if } a \text{ and } b \leq 3 \\ 0 & \text{otherwise,} \end{cases} \tag{21}$$

can be used to generate the (atomic) kinetic temperature upon substitution of eq. 21 into eq. 10:

$$\frac{1}{k_B T_{kin}^A} = \frac{\left\langle \sum_{i\alpha'} \frac{\partial}{\partial \boldsymbol{p}_{i\alpha'}} \cdot \boldsymbol{p}_{i\alpha'} \right\rangle}{\left\langle \sum_{i} \sum_{\alpha'} \frac{\boldsymbol{p}_{i\alpha'} \cdot \boldsymbol{p}_{i\alpha'}}{m_{i\alpha'}} \right\rangle} = \frac{(3Nn_s - K - 3)}{\left\langle \sum_{i} \sum_{\alpha'} \frac{\boldsymbol{p}_{i\alpha'} \cdot \boldsymbol{p}_{i\alpha'}}{m_{i\alpha'}} \right\rangle}, \tag{22}$$

where the second equality follows from taking account of conservation of momentum (removal of 3 degrees of freedom) based on the same reasoning as for the molecular kinetic temperature example, while a further $K$ degrees of freedom are removed because of the holonomic constraints which add a further $K$ relations amongst the momenta (eq. 13b).

### 3. Molecular thermostats for molecules with holonomic constraints

One approach to develop a deterministic thermostat involves the use of Gauss' principle of least constraint. In this method one first derives the form of the equations of motion following which a closed expression is then obtained for the thermostat multiplier. Gaussian isokinetic thermostats are advantageous in that they have no adjustable parameters but have the disadvantage that they do not generate the canonical distribution. This last point is not as serious as might be thought since it has been proved that if the initial distribution is canonical, Gaussian isokinetic dynamics will preserve it [1]. Such thermostats are straightforward to implement for most simple examples[1]. However, fixing the configurational temperatures can be tedious due to the complexity of the constraint multipliers, and if a Gaussian constraint is used to impose the constraint, the canonical ensemble is not preserved [22].

The Nosé-Hoover kinetic thermostat, on the other hand, does generate the canonical distribution of momenta provided that ergodicity holds for the system being studied. An alternative, and extremely useful way to derive the equations of motion for this thermostat was first suggested by Holian in a communication to Bill Hoover[23]. The method, which appears in Hoover's textbook, *Computational Statistical Mechanics*[2], involves solving a generalized Liouville equation in an extended phase space for the canonical distribution function. Knowing the equations of motion for the positions and momenta, this procedure yields the equation of motion for the thermostat degree of freedom.

The DE configurational thermostat was constructed by analogy with the Nosé-Hoover kinetic temperature thermostat in that a term proportional to the gradient of the configurational temperature was added to the position equation of motion in place of a term proportional to the gradient (with respect to momenta) of the kinetic temperature which appears in the momentum equation of motion. It is easy to show that the DE thermostat does not generate the canonical ensemble. The LE thermostat was a generalization of the DE thermostat for molecular systems with constraints[13]. The equations of motion employing this thermostat are given by

$$\dot{\boldsymbol{r}}_{i\alpha} = \frac{\boldsymbol{p}_{i\alpha}}{m_{i\alpha}} - \frac{s}{T_0}\frac{\partial T_{conf}}{\partial \boldsymbol{r}_{i\alpha}} + (\boldsymbol{L}\cdot\varsigma)_{i\alpha} \tag{23}$$

$$\dot{\boldsymbol{p}}_{i\alpha} = \boldsymbol{F}_{i\alpha} + \boldsymbol{F}_{i\alpha}^C \tag{24}$$

$$\dot{s} = -Q\left(\frac{T_{conf} - T_0}{T_0}\right) \tag{25}$$

where $\dot{\mathbf{r}}_{i\alpha}$, $\mathbf{p}_{i\alpha}$ and $\mathbf{F}_{i\alpha}$ are the velocity, momentum and total Newtonian force (arising from inter- and intra-molecular potentials) of site $\alpha$ in molecule $i$, while $\mathbf{F}_{i\alpha}^{C}$ is the force of constraint and $s$ is the thermostat variable. The presence of the 3$^{rd}$ term on the right hand side of eq. 23 is significant. Because the temperature gradient is different at each site in a molecule, the thermostat will work against any holonomic constraint forces resulting in constraint violation. The term, $(\mathbf{L}\cdot\varsigma)_{i\alpha}$, was added by LE to prevent this. The matrix $\mathbf{L}$ is related to the constraint matrix used to define the Gaussian constraints and so is not too difficult to construct, while $\varsigma$ is a new Gaussian multiplier. However, this approach is cumbersome and the LE thermostat still does not generate the canonical distribution.

In a recent publication we discussed the construction of a pair of configurational thermostats that are closer in spirit to the original Nosé-Hoover thermostat in that they generate the canonical ensemble [14]. These thermostats were constructed by following Holian's method described above. The key advantage of these configurational thermostats is the simplicity of the equations of motion; the feedback term is simply proportional to the total force acting on an atom instead of a temperature gradient. We recently learned that a version of the atomic $T_{conF}$ thermostat was also independently derived in Owen Jepps' PhD dissertation[16], which precedes our work.

Encouraged by our initial success we were motivated to generalize our approach to yield a molecular configurational thermostat which a) generates the canonical ensemble, b) has simple feedback terms in the equations of motion and c) does not violate the constraints.

## 3.1 Construction of the $T_{conF}$ thermostat for molecular systems

In developing a molecular configurational thermostat we have chosen to follow the same procedure as we did when developing our configurational thermostats and barostats[14, 24, 25]. However, it is not always obvious what form the equations of motion should have for a given coupling to an external degree of freedom. In the present case we have therefore inverted the Holian-Hoover idea to instead *start* with an expression for the equation of motion of the thermostat degree of freedom, solving a generalized Liouville equation to determine the form of the equation of motion for the positions. This inverse approach is possible because the form of the thermostat variable equation of motion is always known *a priori*. Nosé pointed out that in order to generate the canonical distribution, this feedback is given by the difference of two quantities, the ratio of whose averages in the canonical ensemble is equal to $k_B T$[26].

From the form of eq. 16, a set of motion equations for generating the canonical distribution under configurational temperature control are:

$$\dot{r}_{i\alpha} = \frac{p_{i\alpha}}{m_{i\alpha}} - \zeta \chi_{i\alpha}(r) \tag{26}$$

$$\dot{p}_{i\alpha} = F_{i\alpha} + F_{i\alpha}^C \tag{27}$$

$$\dot{\zeta} = \frac{1}{Q_\zeta} \left( \sum_{i=1}^{N} F_i^2 - k_B T_0 \sum_{i=1}^{N} \sum_{\alpha',\alpha''} \frac{\partial}{\partial r_{i\alpha'}} \cdot \frac{\partial \Phi}{\partial r_{i\alpha''}} \right), \tag{28}$$

where $\zeta$ is the thermostat coupling parameter and $\chi_{i\alpha}(r)$ is the thermostat coupling function. $Q_\zeta$ is an adjustable parameter that may be thought of as a 'mass' associated with the thermostat degree of freedom and is related to the timescale of the temperature feedback and $F_i \left( = \sum_\alpha F_{i\alpha}^N \right)$ is the total force on molecule *i*. The particular form of the coupling function, $\chi_{i\alpha}(r)$, has been restricted to one that is a function of the positions only but within the context of the more general treatment of Kusnezov *et al*[27], it can be a function of positions and momenta. The exact form of $\chi_{i\alpha}(r)$ is determined from the condition

$$\frac{\partial f}{\partial t} + \left( \frac{\partial \dot{r}}{\partial r} + \frac{\partial \dot{p}}{\partial p} + \frac{\partial \dot{\zeta}}{\partial \zeta} \right) f + \dot{r} \frac{\partial f}{\partial r} + \dot{p} \frac{\partial f}{\partial p} + \dot{\zeta} \frac{\partial f}{\partial \zeta} = 0 \tag{29}$$

which is a generalized Liouville equation in the extended phase space, $(\Gamma, \zeta)$. The distribution function, $f$, is related to the canonical distribution function, $f_{NVT}$, by $f = f_{NVT} \times f_\zeta(\zeta)$, where $f_\zeta(\zeta)$ is the distribution function for the thermostat degree of freedom. In writing eq. 29 we have assumed that all the degrees of freedom are independent. For a system of molecules subject to holonomic constraints, the Cartesian positions and momenta of the atoms will not be independent. For a constrained system, the usual procedure in classical mechanics is to construct a Lagrangian using a reduced set of generalized coordinates and their velocities from which a Hamiltonian may be constructed. By adding the constraint relations as additional degrees of freedom, it is possible to construct a phase space of the same dimensions as an unconstrained system involving a set of independent Cartesian coordinates and momenta[28]. The canonical ensemble distribution function for the constrained system is then written over the full phase space with the addition of appropriate delta functions

$$f_{NVT} \propto \exp[-\beta H(\mathbf{r},\mathbf{p})] \prod_{i=1}^{N} \left( |\det \mathbf{Z}(\mathbf{r})| \prod_{k=1}^{s} \delta(\sigma_{ik}) \delta(\dot{\sigma}_{ik}) \right), \tag{30}$$

where $\beta = 1/k_B T$, and $\mathbf{Z}$ is an $s \times s$ matrix with elements given by

$$Z_{i,kj} = \sum_{\alpha=1}^{n_s} \frac{1}{m_{i\alpha}} \left( \frac{\partial \sigma_{ik}}{\partial \mathbf{r}_{i\alpha}} \right) \cdot \left( \frac{\partial \sigma_{ij}}{\partial \mathbf{r}_{i\alpha}} \right), \qquad j, k = 1, s \tag{31}$$

with a Hamiltonian defined by

$$H(\mathbf{r},\mathbf{p}) = \sum_i \sum_\alpha \frac{p_{i\alpha}^2}{2m_{i\alpha}} + \Phi(\mathbf{r}) \tag{32}$$

However, unlike the atomic case, the state points are restricted to a phase space hypersurface upon which all constraints defined by (eqs. 13a – 13b), are satisfied. The flow of phase points in the Liouville equation is guaranteed by the use of Gauss' principle of least constraint in the derivation of the constrained equations of motion.. The non-vanishing terms in the Liouville equation are

$$\left(\frac{\partial}{\partial \mathbf{r}} \cdot \dot{\mathbf{r}}\right) f = -\zeta f \sum_i \sum_\alpha \frac{\partial}{\partial \mathbf{r}_{i\alpha}} \cdot \mathbf{\chi}_{i\alpha}(\mathbf{r}) \qquad (33)$$

$$\dot{\mathbf{p}} \cdot \frac{\partial f}{\partial \mathbf{p}} = -\beta f \sum_i \sum_\alpha \frac{\mathbf{p}_{i\alpha}}{m_{i\alpha}} \cdot \left(\mathbf{F}_{i\alpha} + \mathbf{F}_{i\alpha}^C\right) \qquad (34)$$

$$\dot{\mathbf{r}} \cdot \frac{\partial f}{\partial \mathbf{r}} = \beta f \sum_i \sum_\alpha \left(\frac{\mathbf{p}_{i\alpha}}{m_{i\alpha}} - \zeta \mathbf{\chi}_{i\alpha}(\mathbf{r})\right) \cdot \mathbf{F}_{i\alpha} \qquad (35)$$

$$\dot{\zeta} \frac{\partial f}{\partial \zeta} = \frac{1}{Q_\zeta}\left(\partial \ln f_\zeta(\zeta)/\partial \zeta\right) f \left(\sum_i \mathbf{F}_i^2 + kT_0 \sum_i \sum_\alpha \frac{\partial}{\partial \mathbf{r}_{i\alpha}} \cdot \mathbf{F}_i\right) \qquad (36)$$

Substituting the above equations into eq. 29 and simplifying, gives

$$\beta \sum_i \sum_\alpha \mathbf{\chi}_{i\alpha} \cdot \mathbf{F}_{i\alpha} + \sum_i \sum_\alpha \frac{\partial}{\partial \mathbf{r}_{i\alpha}} \cdot \mathbf{\chi}_{i\alpha} = \beta \sum_i \mathbf{F}_i^2 + \sum_i \sum_\alpha \frac{\partial}{\partial \mathbf{r}_{i\alpha}} \cdot \mathbf{F}_{i\alpha}, \qquad (37)$$

where we have used the identity,

$$Q_\zeta = \frac{1}{\zeta}\left(\partial \ln f_\zeta(\zeta)/\partial \zeta\right) \qquad (38)$$

Comparing terms on both sides of eq. 37 leads to the identification

$$\mathbf{\chi}_{i\alpha} = \sum_\alpha \mathbf{F}_{i\alpha} \equiv \mathbf{F}_i \qquad (39)$$

Eq. 39 shows that the feedback term is proportional to the net force acting on the molecule. This is particularly advantageous because the feedback will be identical for all sites in a molecule and hence there will be no constraint violation as a result of the thermostat.

The final form for the molecular $T_{conF}$ equations of motion is therefore

$$\dot{r}_{i\alpha} = \frac{p_{i\alpha}}{m_{i\alpha}} + \zeta \, F_i \tag{40}$$

$$\dot{p}_{i\alpha} = F_{i\alpha} + F_{i\alpha}^{C} \tag{41}$$

$$\dot{\zeta} = \frac{1}{Q_\zeta}\left( \sum_i F_i^2 + k_B T_0 \sum_i \sum_\alpha \frac{\partial}{\partial r_{i\alpha}} \cdot F_i \right) \tag{42}$$

The above form of the equations of motion makes this algorithm particularly simple to implement in a standard Molecular Dynamics code. Unlike the LE thermostat, the instantaneous configurational temperature is not required. The squared force term involves a simple order *N* loop, while the configurational Laplacian term is evaluated inside the *intermolecular* force double loop. There are no contributions to this latter term from intramolecular degrees of freedom such as torsional motion because the forces arising from them sum to zero for each molecule. Aside from a trivial change to the squared force calculation, the code for the molecular configurational thermostat algorithm is very similar to that used for the atomic version of this thermostat. The LE configurational thermostat on the other hand, requires extra blocks of code for the calculation of the contribution made by these intramolecular degrees of freedom to the site based gradient of the configurational temperature.

If the potential energy of a system of molecules is pairwise additive and given by the Lennard-Jones 12-6 potential,

$$\Phi = \frac{1}{2}\sum_i \sum_{\alpha'} \sum_j \sum_{\alpha''} U^{LJ}\left(\left|r_{i\alpha'j\alpha''}\right|\right) \tag{43}$$

where

$$U^{LJ} = 4\varepsilon_{i\alpha'j\alpha''}\left\{\left(\frac{\sigma_{i\alpha'j\alpha''}}{\left|r_{i\alpha'j\alpha''}\right|}\right)^{12} - \left(\frac{\sigma_{i\alpha'j\alpha''}}{\left|r_{i\alpha'j\alpha''}\right|}\right)^{6}\right\} \tag{44}$$

the gradient term in the configurational Laplacian term is given by

$$\frac{\partial}{\partial r_{i\alpha'j\alpha''}} \cdot \frac{\partial U^{LJ}}{\partial r_{i\alpha'j\alpha''}} = \frac{-528\varepsilon_{i\alpha'j\alpha''}}{\left|r_{i\alpha'j\alpha''}\right|^2}\left\{\left(\frac{\sigma_{i\alpha'j\alpha''}}{\left|r_{i\alpha'j\alpha''}\right|}\right)^{12} - \frac{5}{22}\left(\frac{\sigma_{i\alpha'j\alpha''}}{\left|r_{i\alpha'j\alpha''}\right|}\right)^{6}\right\} \tag{45}$$

The configurational Laplacian for non-Lennard-Jones potentials may be more complicated or may not even exist in an analytical form. However, a simple workaround for these cases is to estimate it using finite differences[25]. This will be particularly useful for general simulation codes which include support for many different potential functions.

Finally we give the equations of motion for controlling the order 1 configurational temperature, $T_{con1}$. These are derived in the same way as above except that the thermostat variable equation of motion is constructed from the definition of temperature given by eq. 15:

$$\dot{r}_{i\alpha} = \frac{p_{i\alpha}}{m_{i\alpha}} + \frac{\zeta}{\Delta}F_i \tag{46}$$

$$\dot{p}_{i\alpha} = F_{i\alpha} + F_{i\alpha}^C \tag{47}$$

$$\dot{\zeta} = \frac{1}{Q_\zeta}\left(\frac{\sum_i F_i^2}{\Delta} - k_B T_0\right) \qquad (48)$$

where $\Delta$ is the configurational Laplacian, $\sum_i \sum_\alpha \frac{\partial}{\partial r_{i\alpha}} \cdot F_i$. While both of the above configurational thermostats employ centre-of-mass based feedback in the equations of motion, they are not 'molecular' in the same sense as the molecular based kinetic thermostat since in the latter case, the thermostat is inhomogeneous, operating on only 3 translational degrees of freedom. Our two configurational thermostats are closer in spirit to the atomic kinetic thermostat.

4. **Validation of the new molecular configurational thermostat**

Any new algorithm proposed for use in Molecular Dynamics should be thoroughly tested/validated against known simulation data. Our new configurational thermostat is no exception and we have therefore chosen to validate it by conducting equilibrium MD simulations of a fluid composed of semi-flexible *n*-decane molecules. This system was chosen because the model *n*-decane (described in the next section) has a sufficient degree of intramolecular flexibility and constrained bond lengths and angles to be a reasonable test for a configurational thermostat. This model has been studied previously at the same state conditions.

4. 1 **Molecular Model**

The model used for *n*-decane is based on the one introduced by Ryckaert and Bellemans[29]. The model *n*-decane molecule is therefore constructed from 10 equivalent united atom sites representing methyl and methylene groups. Each of these sites has a mass equal to 1/10 of the mass of an *n*-decane molecule. The distance between adjacent sites (bond length) are fixed at 1.53 Å, while valence bond angles are fixed at 109.47˚. Torsional motion within the molecules is controlled by a torsional potential of the form

$$U_{tors} = \sum_{\alpha}^{n_d} \sum_{n=0}^{5} C_n (\cos\phi_\alpha)^n \tag{49}$$

where $n_d$ is the number of dihedral angles in the molecule (= 7 in the united atom model of *n*-decane), $\phi_\alpha$ is the $\alpha$th dihedral angle in a given molecule, and $C_n$ are potential parameters given by $C_n/R = [1.116, 1.462, -1.578, -0.368, 3.156, -3.788\}] \times 10^3$ K, where $R$ is the molar gas constant.

United atom sites on different molecules and sites more than three sites apart on the same molecule interact through a WCA potential (a Lennard-Jones potential truncated and shifted at its minimum, $r = 2^{1/6} \sigma$). The parameters used in the WCA potential were $\sigma = 3.923$ Å, $\varepsilon/k_B = 72$ K.

### 4.2 Simulation Details

Gaussian constraints were employed to fix the nearest, and second nearest neighbor distances in *n*-decane. These were used in conjunction with feedback terms to cancel any numerical drift in the values of the constrained distances. Full details on how to implement Gaussian bond constraints and feedback can be found in the articles by Morriss and Evans[30] and Daivis *et al*[31].

All simulations were conducted at the same reduced temperature, $T^* \equiv Tk_B/\varepsilon = 5$, and reduced site number density, $\rho_s^* (\equiv \rho_s \sigma^3) = 1.537$ using a system size of 256 molecules.

The appropriate equations of motion (using either kinetic or configurational thermostats) were integrated using a 5$^{th}$ order Gear algorithm with a reduced time step $\tau^* \equiv \tau\left(\sigma\sqrt{(M_i/\varepsilon)}\right)^{-1} = 0.0001$.

The kinetic temperature Nosé-Hoover equations of motion are

$$\dot{r}_{i\alpha} = \frac{p_{i\alpha}}{m_{i\alpha}} \tag{50}$$

$$\dot{p}_{i\alpha} = F_{i\alpha} + F_{i\alpha}^C - \xi \frac{m_{i\alpha}}{M_i} P_i \tag{51}$$

$$\dot{\xi} = \frac{1}{Q_\xi}\left(\sum_i \mathbf{P}_i \cdot \mathbf{P}_i / M_i - 3(N-1)k_B T_0\right) \quad (52)$$

where $\xi$ and $Q_\xi$ are the kinetic thermostat variable and mass term. The configurational thermostat equations of motion are given by eqs. 40-42.

Since all our simulations were conducted using reduced units we shall henceforth report all quantities in this dimensionless form, dropping the asterisk notation unless there is a potential for ambiguity.

Starting configurations for the simulations were obtained by first constructing a set of *fcc* lattice positions with the appropriate molecular density and then placing *n*-decane molecules such that their centres of mass coincided with these lattice positions. The sites in each molecule were constructed according to an all-*trans* conformational state for simplicity and each molecule was given a random orientation. This configuration was then melted at $T = 5$ using molecular dynamics with the truncated force method [32] and kinetic temperature thermostatting. This phase lasted for 500,000 timesteps. A further 100,000 steps were performed using the un-modified WCA potential until the translational and configurational temperatures agreed with each other. This configuration was then used as a starting state for production runs comprising of 1 million steps.

The thermostat mass (inertia) factors used with the kinetic and configurational Nosé-Hoover thermostats were: $Q_\xi = 100$, and $Q_\zeta = 10^7$, respectively.

## 5. Results

The averages taken from the production phase simulations involving both configurational ($\text{NVT}_{\text{conF}}$) and kinetic ($\text{NVT}_{\text{kin}}$) based thermostats are collected in Table 1. Within the statistical uncertainty, all calculated properties agree between the two sets of simulation results. Both thermostats have successfully maintained their respective temperatures at the set point value and the various measures of temperature are in agreement with one notable exception – under $NVT_{conF}$ dynamics, $<T_{con1}>$ is lower than $<T_{conF}>$. It is well known that $T_{conF}$ is a more accurate measure of configurational temperature than $T_{con1}$, but on that basis we would expect to observe a similar disparity between these two temperatures under $NVT_{kin}$ dynamics. Our results do not bear this out. However, the statistical uncertainty is much larger in the value obtained with this thermostat, and may conceal any real differences. More precise simulation data could be obtained to investigate this further, but we feel this would be nothing more than an interesting academic exercise with no obvious reason for preferring $T_{con1}$ over $T_{conF}$. It should be noted that both the DE and LE configurational thermostats are closest in form respectively to the $NVT_{con1}$ thermostat described in our previous paper[14], and the molecular one in this work.

The structure of the *n*-decane molecules is clearly unaffected by the molecular configurational thermostat. Two measures of the length of the molecules: the root mean square (rms) radius of gyration, $\left\langle R_g^2 \right\rangle^{1/2}$, and rms end-to-end distance, $\left\langle \Gamma_{1n}^2 \right\rangle^{1/2}$, are both in agreement between both sets of simulation data. A more stringent validation is provided by the distribution of these quantities. The normalized distribution of rms end-to-end distances obtained under $NVT_{kin}$ and $NVT_{conF}$ dynamics are plotted in Figure 1. There is a perfect match between the two sets of data points. The intramolecular structure may also be examined by calculating the distribution of torsion angles. Figure 2 shows this distribution calculated under both sets of dynamics. Again, excellent agreement is found between the two sets of data.

The coefficient of self-diffusion, $D_S$, has been calculated using the Einstein mean square displacement method. Table 1 shows good agreement between the diffusion coefficients calculated by the two simulation methods. Stronger proof that dynamical properties are unaffected by the type of thermostat is provided by Figures 3 and 4. Figure 3 shows the actual mean square displacements versus time while Figure 4 shows the short time behavior of the (atomic) velocity autocorrelation function. In both cases there is excellent agreement between the two sets of data. We therefore conclude that under equilibrium conditions, the molecular configurational thermostat gives results in agreement with those obtained using the molecular kinetic thermostat.

## 6. Discussion and Conclusion

We have derived a new configurational temperature thermostat suitable for molecules with holonomic constraints. This thermostat has been derived using a method similar to the one we used to derive atomic configurational thermostats and barostats in our previous papers[14 24 25]. The method entails solving a generalized Liouville equation in an extended phase space in order to determine the form of a generalized force that is coupled to the position equation of motion. This is subtly different to the approach we have used previously, where in that case, the coupling force was known *a priori* and the solution of Liouville's equation led to the equation of motion for the thermostat/barostat variable. The present "inverse" approach makes use of Nosé's ansatz that the equation of motion for the thermostat variable will be proportional to the difference of two terms, the ratio of whose averages in the canonical ensemble is equal to $k_BT$. This approach could be very useful for developing other, more general sets of motion equations in the spirit of Kusnezov *et al*[27]. Following this procedure, working backwards from the expression for the configurational temperature suitable for constrained molecules, we determined that the generalized force acting through the position equation of motion is proportional to the net force acting on a molecule. This result is significant because it leads to a molecular configurational thermostat that does not intrinsically violate the constraints – unlike the only other known molecular configurational thermostat, first proposed by Lue and Evans. The equations of motion do not involve any gradients of the configurational temperature, nor do they require knowledge of the instantaneous configurational temperature itself.

Other key features of our configurational thermostat are that it will generate the canonical distribution in both position and momentum space and that it acts homogeneously through the atomic coordinates and thus is the configurational analogue of the atomic based kinetic thermostat (unlike the translational thermostat which operates on only 3 degrees of freedom).

Our new thermostat is very straightforward to implement in a typical Molecular Dynamics computer code, requiring trivial changes to the coding of an atomic configurational thermostat. Because the thermostat feedback and configurational temperature expression both involve the net molecular force, contributions from intramolecular degrees of freedom do not need to be taken into account (they cancel exactly).

We have validated our new thermostat by conducting equilibrium molecular dynamics simulations using a well-known semi-flexible model of *n*-decane, and comparing static and dynamic properties against those calculated under kinetic Nosé-Hoover dynamics. Given the fact that both thermostats yield similar results, we now attempt to answer the question of when and where to use configurational thermostats.

The first point that needs to be made (or reinforced) is that eq. 1 is an equilibrium expression and thus Rugh's/Jepps *et al*/Rickayzen and Powles' expressions for temperature are derived with the implicit assumption that the systems remain in thermal equilibrium. At equilibrium, all the expressions are valid and so one is free to choose the simplest and most convenient definition of temperature. Almost always this will be the kinetic temperature. However, there are cases in which it may be difficult to control the kinetic temperature. Two examples, though not strictly equilibrium systems, are nevertheless relevant to this discussion. These are: temperature quench molecular dynamics, and radiation damage modeling. In the former example, heat can be removed at a very high rate by a thermostat while in the second example, a large quantity of heat is instantaneously introduced (via a single particle) into a solid and this is then allowed to dissipate usually in the presence of some thermostatting boundary layer. In both of these cases, control of the configurational temperature can avoid the difficulty associated with the speed by which energy can be removed or imparted on the kinetic degrees of freedom. In another example involving the simulation of large molecules, it may prove more efficient to guide the simulation to equilibrium by controlling the configurational temperature since the relaxation times for the many intramolecular degrees of freedom of a large molecule such as a polymer can be disparate. All the aforementioned examples can be expected to lead to new and interesting research directions.

We now turn to non-equilibrium simulations. Here, we have the basic difficulty whereby it is not at all clear which definition of temperature (if any) can be identified with the so-called non-equilibrium temperature. At equilibrium, the zeroth law of thermodynamics permits the existence of an empirical temperature. However, there is no zeroth law in a non-equilibrium situation [33]. The postulate of local thermodynamic equilibrium allows us to extend the definitions of variables such as temperature and entropy away from equilibrium and we are thus permitted to used eq. 1 in the weak field regime, at least. At higher fields, the issue becomes much thornier; some kind of thermostat (or ergostat) is necessary in order to obtain a non-equilibrium steady state, but which temperature is chosen as the operational definition remains an open question. Usually one takes a pragmatic approach and chooses the kinetic temperature; this being a convenient, intuitive and simple object to determine. However, there is no theoretical proof that the kinetic temperature is equivalent to the non-equilibrium temperature. When, and why would one choose to thermostat the configurational temperature in a non-equilibrium simulation, then? An interesting first example is the case of thermostatting a system of large molecules with many degrees of freedom under some non-equilibrium flow. Thermostatting the total kinetic energy would be desirable for the homogeneous removal of heat. However, such a thermostat requires an *a priori* knowledge of the streaming velocity at an atomic site. In the weak field regime, the translational streaming velocity is usually known. The angular streaming velocity, by contrast, is not. An incorrect assumed form for this streaming velocity will lead to the imposition of spurious torques, resulting in an antisymmetric stress tensor. This effect is best illustrated in the work of Travis *et al* [12] who simulated a system of linear rigid rods undergoing planar Couette flow. In this example it is clear that a configurational thermostat (which makes no assumptions about the streaming velocity) is to be preferred.

Outside the linear regime, the above mentioned problems related to knowing the instantaneous local streaming velocity make kinetic thermostatting even more problematic. Configurational temperature control appears more appealing under these circumstances. In lieu of any proof via theory or simulation that configurational temperature is a valid definition of temperature very far from equilibrium, it may be more fruitful to develop better kinetic thermostats, perhaps operating on higher moments of the velocity distribution, or using velocity differences as per DPD.

Finally we note that theoretical developments concerning the non-equilibrium temperature constitute an highly active field of research. For an excellent overview of the state of the art, see the recent review on this topic by Casas-Vázquez and Jou. One of the first steps towards developing a non-equilibrium temperature was taken by Morriss and Rondoni [34]. These authors studied the thermostatted color diffusion of soft disks, a system which is Hamiltonian. They studied the effect of controlling temperature via the momenta, the coordinates, and a weighted mixture of the two, finding a correlation between the orthogonal (to the flow) component of kinetic temperature and the thermodynamic temperature defined by Rugh.

ACKNOWLEDGMENT.

The authors thank EPSRC for provision of funding through Grant Number GR/R13265/01), and Bill Hoover for some helpful and stimulating discussions during the preparation of the manuscript.

FIGURE CAPTIONS

**Figure 1**. Plot of root mean square end-to-end distribution obtained under $NVT_{conF}$ dynamics (solid line) and $NVT_{kin}$ dynamics (filled circles).

**Figure 2**. Plot of torsion angle distribution obtained under $NVT_{conF}$ dynamics (solid line) and $NVT_{kin}$ dynamics (filled circles).

**Figure 3**. Plot of (COM) mean square displacements obtained under $NVT_{conF}$ dynamics (solid line) and $NVT_{kin}$ dynamics (filled circles).

**Figure 4**. Plot of the short time part of the (atomic) velocity autocorrelation function obtained under $NVT_{conF}$ dynamics (solid line) and $NVT_{kin}$ dynamics (filled circles).

TABLES.

**Table 1.** - Equilibrium results for *n*-decane for the state point: $T = 5.0$, $\rho_s = 1.537$, obtained using both $T_{kin}$NH and $T_{conF}$NH thermostats. Numbers in parentheses give the statistical uncertainty in the last digit.

| Property | $T_{kin}$NH | $T_{conF}$NH |
|---|---|---|
| $\langle T_{kin}^A \rangle$ | 5.002(5) | 5.002(6) |
| $\langle T_{kin}^M \rangle$ | 5.000(1) | 4.997(9) |
| $\langle T_{con1} \rangle$ | 5.003(7) | 4.990(3) |
| $\langle T_{conF} \rangle$ | 5.003(7) | 5.000(3) |
| $\langle p \rangle$ | 11.84(2) | 11.81(2) |
| $\langle U^{LJ} \rangle$ | 6.138(7) | 6.136(7) |
| $\langle U^{intra\,LJ} \rangle$ | 0.929(6) | 0.928(6) |
| $\langle U^{tor} \rangle$ | 36.11(5) | 36.03(5) |
| $\langle \%trans \rangle$ | 64.63(6) | 64.67(7) |
| $\langle R_g^2 \rangle^{1/2}$ | 0.7993(2) | 0.7995(2) |
| $\langle \Gamma_{1n}^2 \rangle^{1/2}$ | 2.290(1) | 2.292(1) |
| $D_S$ | 0.251(5) | 0.245(5) |

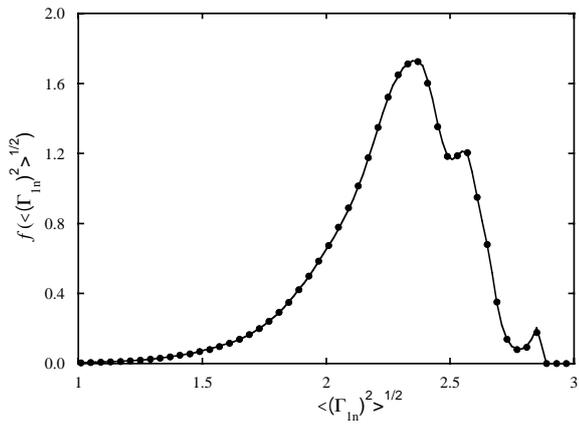

Figure 1

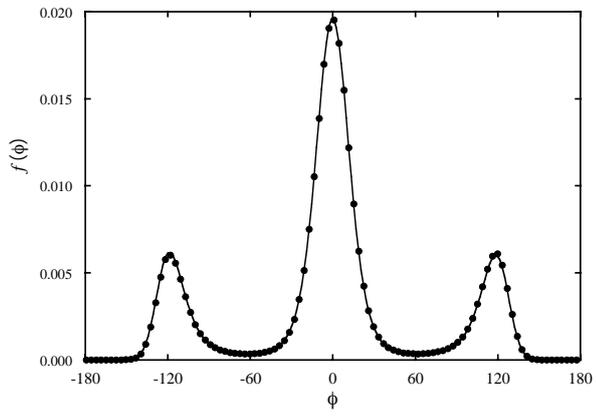

Figure 2

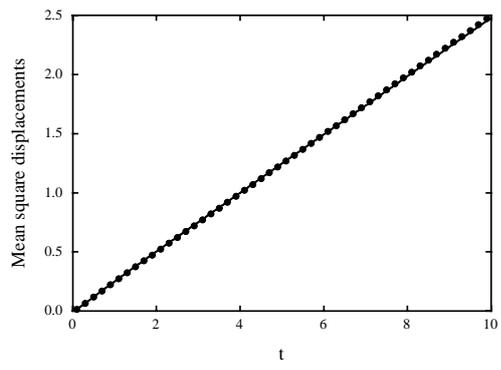

Figure 3

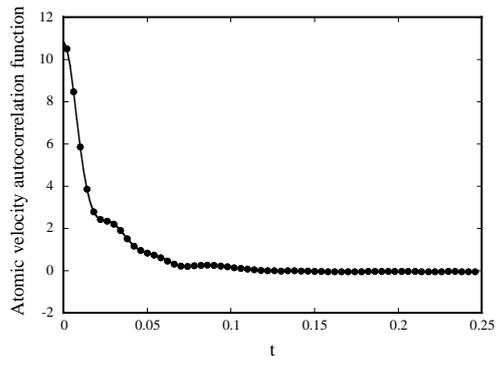

Figure 4